\documentclass{elsart}

 \usepackage{epsfig}

\usepackage{amssymb}
\begin{document}
\begin{frontmatter}


\title{Metastability
and anomalous behavior in the HMF Model: connections to
nonextensive thermodynamics and glassy dynamics   }

\author
{Alessandro Pluchino},
\author
{Andrea Rapisarda}, 
\author
{Vito Latora}   

\address{Dipartimento di Fisica e Astronomia,  Universit\'a di Catania,\\
and INFN sezione di Catania, Via S. Sofia 64,  I-95123 Catania, Italy}

\begin{abstract}

We review some of the most recent results on the dynamics
of the Hamiltonian Mean Field (HMF) model, a systems of N planar spins
with ferromagnetic infinite-range interactions.
We show, in particular, how some of the dynamical anomalies of the
model can be interpreted and characterized in terms of the weak-ergodicity
breaking proposed in frameworks of glassy systems. We also discuss the
connections with the nonextensive thermodynamics proposed by
Tsallis.

\end{abstract}

\end{frontmatter}

\section{{Introduction}}

Metastability, nonextensivity and glassy dynamics are features
so ubiquitous in complex systems that are often used to
characterize them or, more in general, to define \textit{complexity}
\cite{yaneer}.
In this work we will discuss such features in the context of the
so called \textit{Hamiltonian Mean Field} (HMF) model, a
system of inertial spins with long-range interaction.
The model, originally introduced by Antoni and Ruffo \cite{hmf0},
has been thoroughly investigated and generalized in the last
years for its anomalous dynamical behavior
\cite{hmf1,hmf-dif,hmf2,hmf3,alfaxy,plu1,celia1,plu2,plu3,yama,chava}.
With respect to systems with short-range interactions,
the dynamics and the thermodynamics of many-body systems
of particles interacting with long-range forces,
as the HMF, are particularly rich and interesting.
In fact, more and more frequently nowadays,
the out-of-equilibrium dynamics of systems with  long-range
interactions or long-term correlations
has shown physical situations which can be badly 
described within the ergodic assumption that is at the basis of
the Boltzmann-Gibbs thermostatistics. 
In all such cases it happens, for instance, that a
system of particles kept at constant total energy $E$, does not
not visit all the a-priori available phase space (the surface
of constant energy E), but it seems to remain trapped in
a restricted portion of that space, giving rise to anomalous
distributions that differs from those expected.
A few years ago, Tsallis has introduced
a generalized thermodynamics formalism based on a
nonextensive definition of entropy \cite{tsallis0}.
This nonextensive thermodynamics is very  useful in describing 
all those situations characterized by long-range correlations
or fractal structures in phase space \cite{hmf3,tsallis1,tsallis-last}.
On the other hand, the latter feature is also connected with the
so called "weak ergodicity breaking" scenario, which is at the
basis of the long-term relaxation and aging observed in glassy
systems. Such systems show competing interactions
({\it frustration}) and are characterized by a complex landscape
and a hierarchical topology in some high dimensional configuration
space \cite{spin-glass}, which, in turn, generates a strong
increase of  relaxation times together with 
metastable states and weak chaos.
\\
The Hamiltonian Mean Field model, considered in this paper,
is exactly solvable at equilibrium and exhibits a series
of anomalies in the dynamics, as the presence
of quasistationary states (QSS) characterized by:
anomalous diffusion, vanishing Lyapunov exponents,
non-gaussian velocity distributions, aging and  fractal-like
phase space structure.
Furthermore, the model is easily accessible by means
of both molecular dynamics and Monte Carlo simulations.
Thus, it represents a very useful ``laboratory'' for exploring
metastability and glassy dynamics in systems with long-range
interactions.
The model can be considered as a  minimal  and pedagogical model
for a large class of complex systems,
among which one can surely include self-gravitating systems \cite{chava}
and glassy systems \cite{plu2}, but also systems apparently
belonging to different fields as biology or sociology.
In fact, we recently found similar features also in the context of
the \textit{Kuramoto Model} \cite{kuramoto}, one of the
simplest models for synchronization in biological
systems \cite{kurastab}. Moreover,
the proliferation of metastable states in the vicinity of
a critical critical point in the phase diagram seems to be
responsible for the onset of complexity and diverging time
calculation in many different kind of algorithms \cite{parisi-science}.
\\
In this paper we focus on two different aspects of the HMF model:
its glassy-dynamics and the possible connections with the generalized
thermodynamics. The paper is divided into two parts.
In Section 2.1 we investigate the model
following the analogy with glassy systems
and the 'weak ergodicity breaking' scenario. In
previous works we have shown that the ``thermal explosion'',
characteristic of initial conditions with  finite magnetization,
drives the system into a metastable glassy-like regime which
exhibits 'dynamical frustration'. With the aim to characterize in
a quantitative way  this behavior, we have explicitly suggested to
introduce a new order parameter, the 'polarization', able to
measure the degree of freezing of the rotators (or particles).
Here we present new numerical results reinforcing the glassy
nature of the QSS's metastability and the hierarchical
organization of phase space.
In Section 2.2 we investigate the links with
 nonextensive thermostatistics. In ref.\cite{plu3} we have
found that, for a particular class of initial conditions with constant
velocity distribution and finite magnetization,
the velocity correlations obtained by integrating the equations
of motion of the HMF model are well reproduced by q-exponential curves.
Here, we show numerical evidences that the superdiffusion observed
in the anomalous QSS regime (ref. \cite{hmf2}) can be linked
with the q-exponential long-term decay of the velocity correlations,
as analitically suggested by a formula obtained by Tsallis
and Bukman \cite{tsa-buk} for a {\it nonlinear} Fokker-Planck (FP) 
equation, using an ansatz based on the generalized entropy.

\section{{Anomalous dynamics in the HMF model}}

The HMF model has been introduced originally in ref.\cite{hmf0}
with the aim of studying {\it clustering} phenomena in $N$-body
systems in one dimension. The Hamiltonian of the ferromagnetic HMF
model is:
\begin{equation}
H = K + V = \sum_{i=1}^N \frac{p_i^2}{2} + \frac{1}{2N}
\sum_{i,j=1}^N [1-\cos(\theta_i - \theta_j)]~~~~,
\label{model0}
\end{equation}
\noindent where the potential energy is rescaled by $1/N$ in order
to get a finite specific energy in the thermodynamic limit $N
\rightarrow \infty$. This model can be seen as classical
$XY$-spins (inertial rotators) with unitary masses and infinite
range coupling, but it also represents particles moving on the
unit circle. In the latter interpretation the coordinate
$\theta_i$ of particle $i$ is its position on the circle and $p_i$
its conjugate momentum (or velocity). Associating to each particle
the spin vector
\begin{displaymath}
\overrightarrow{s}_i =(\cos \theta_i, \sin \theta_i)~~,
\end{displaymath}
it is possible to introduce the
following mean-field order parameter:
\begin{equation}
M = \frac{1}{N} |\sum_{i=1}^N \overrightarrow{s}_i |~~,
\label{m0}
\end{equation}
representing the modulus of the total {\em magnetization}.
\\
The equilibrium thermodynamical solution in the canonical
ensemble predicts a
second-order phase transition from a low-energy condensed
(ferromagnetic) phase with magnetization  $M\ne0$, to a
high-energy one (paramagnetic), where the spins are homogeneously
oriented on the unit circle and $M=0$. The {\em caloric curve},
i.e. the dependence of the energy density $U = H/N$ on the
temperature $T$, is given by
$U = \frac{T}{2} + \frac{1}{2} \left( 1 - M^2 \right)
~$\cite{hmf0,hmf2}.
The critical point is at energy density $U_c=\frac{3}{4}$,
corresponding to a critical temperature $T_c=\frac{1}{2}$.
\\
At variance with the equilibrium scenario, the out-of-equilibrium
dynamics shows, just below the phase transition, several anomalies
before complete equilibration. More precisely, if we adopt the
so-called $M1$ initial conditions (i.c.), i.e. $\theta_i=0$  for
all $i$ ($M(0)=1$) and velocities uniformly distributed ({\it
water bag}), the results of the simulations, in a special region
of energy values (in particular for $0.68<U<U_c$) show a disagreement with
the canonical prediction for a transient regime whose length
depends on the system size N. In such a regime, the system remains
trapped in metastable states (QSS) with vanishing magnetization
at a temperature lower then the
canonical equilibrium one, until it slowly relaxes towards
Boltzmann-Gibbs (BG)  equilibrium, showing strong memory effects,
correlations and aging. This transient QSS regime becomes stable
if one takes the infinite size limit before the infinite time
limit.

\subsection{{Dynamical frustration and hierarchical structure}}

\begin{figure}
\label{fig1}
\begin{center}
\epsfig{figure=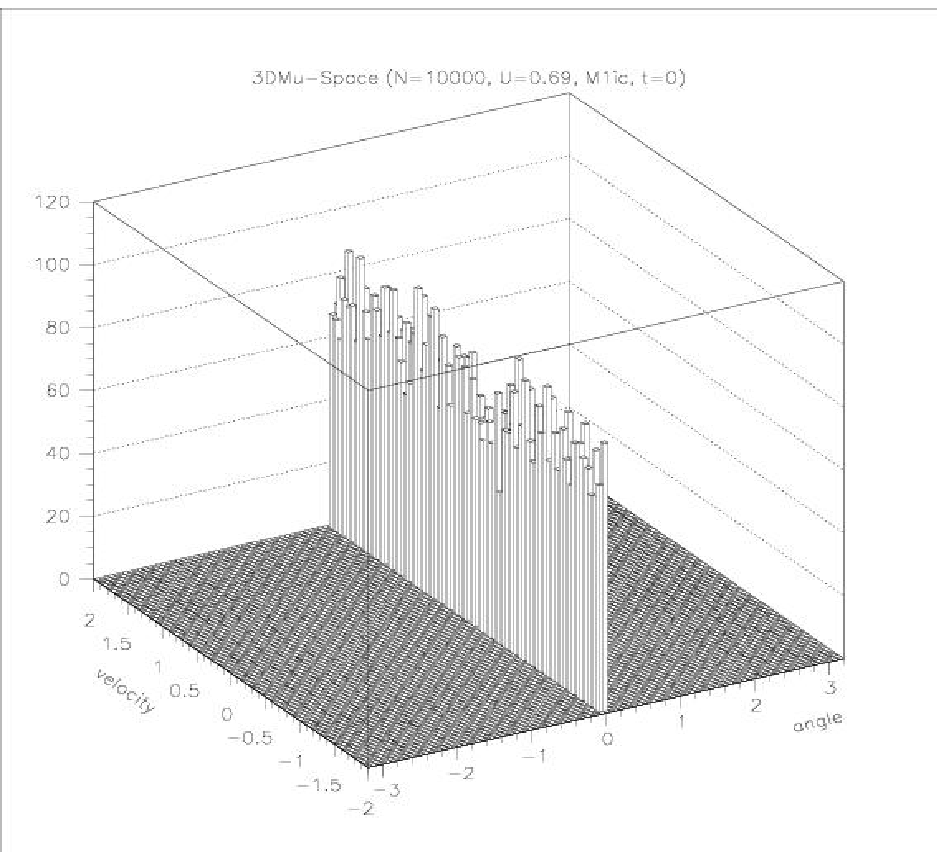,width=6truecm,angle=0}
\epsfig{figure=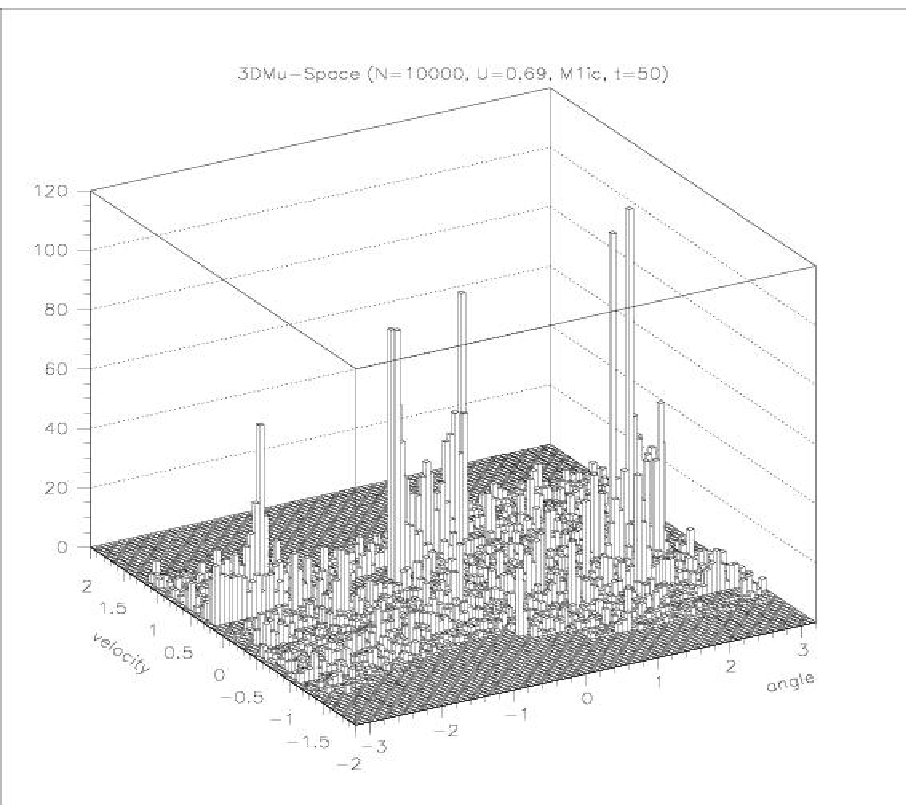,width=6truecm,angle=0}
\end{center}
\begin{center}
\epsfig{figure=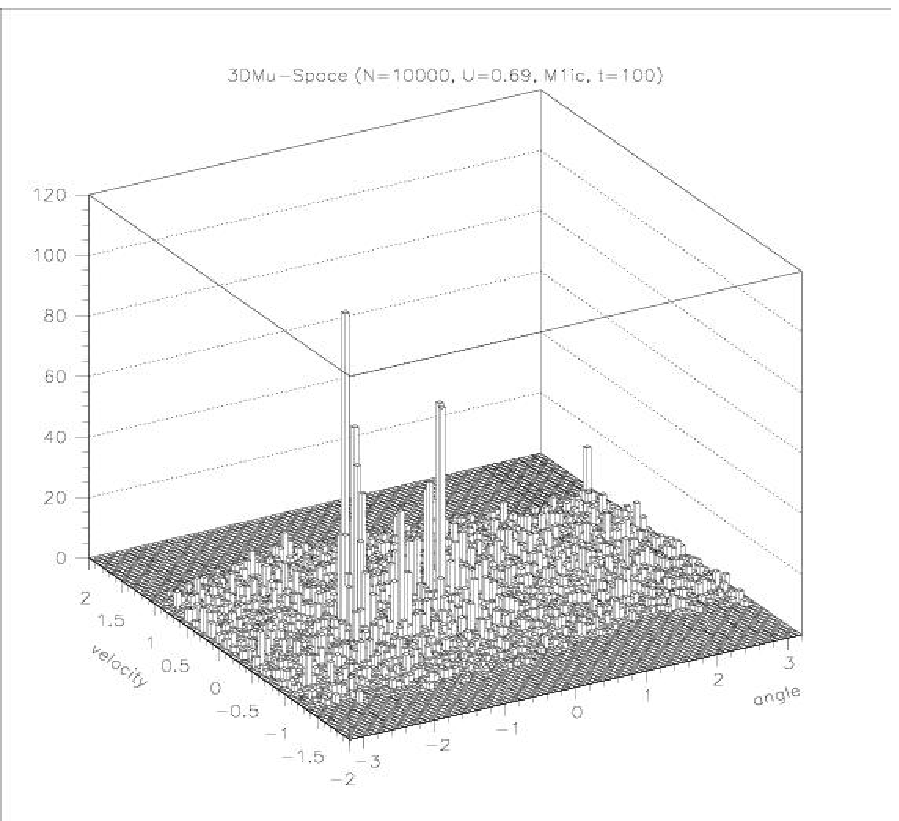,width=6truecm,angle=0}
\epsfig{figure=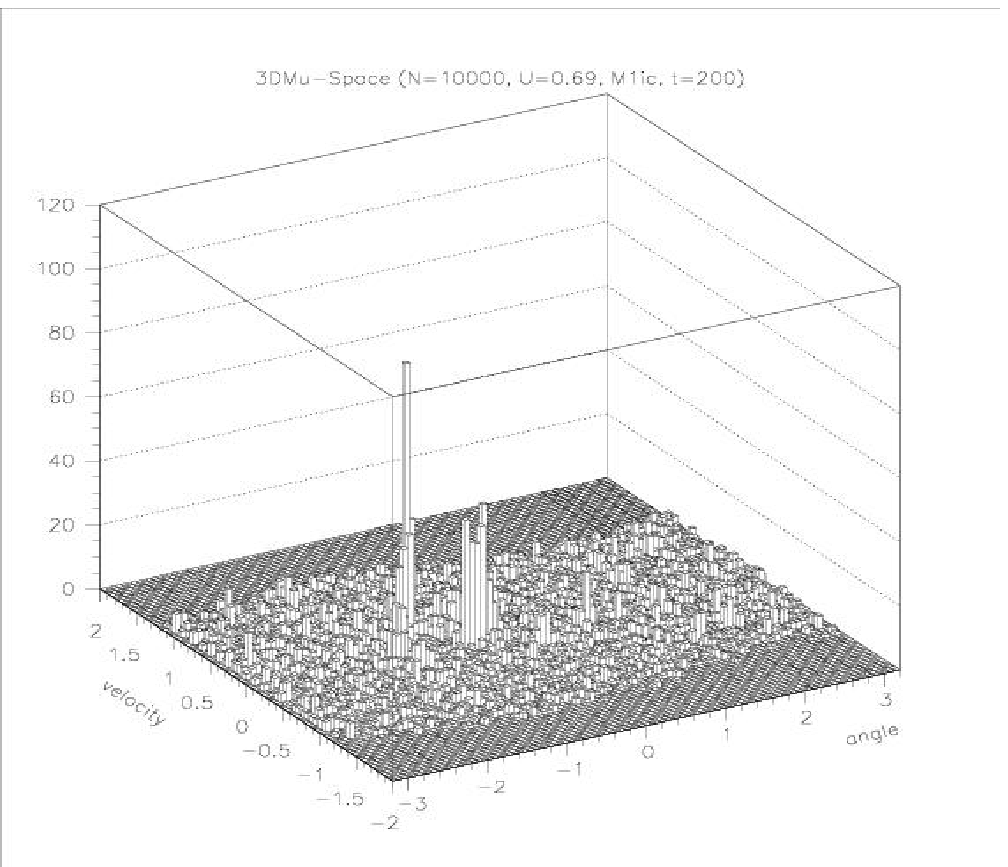,width=6truecm,angle=0}
\end{center}
\begin{center}
\epsfig{figure=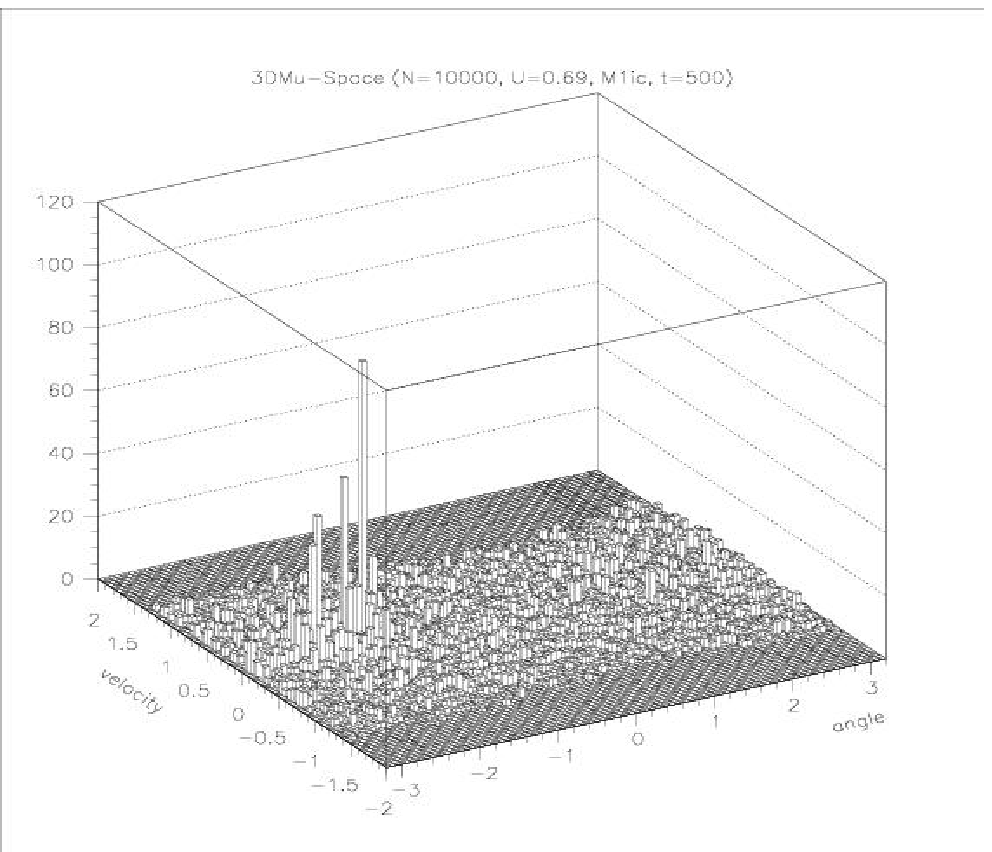,width=6truecm,angle=0}
\epsfig{figure=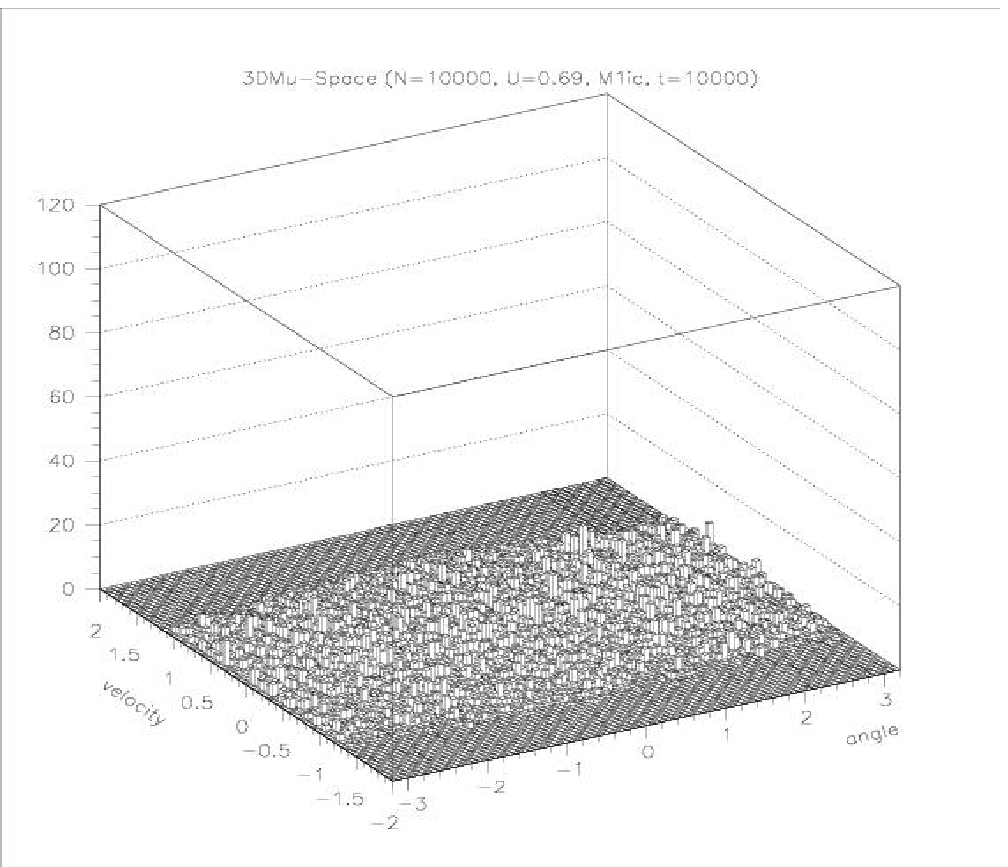,width=6truecm,angle=0}
\end{center}
\caption{Snapshots of $f(\theta, p, t)$ for t=0 (upper left), 50
(upper right), 100 (center left), 200 (center right), 500 (lower
left) and 1000000 (lower right). In this case we considered
N=10000 at U=0.69 and M1 i.c.. See text. 
 }
\end{figure}

\begin{figure} 
\begin{center}
\epsfig{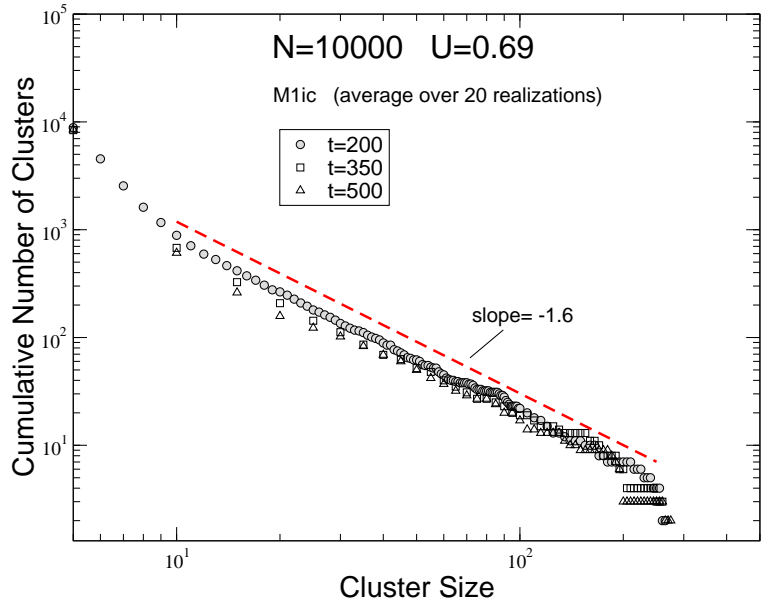}
\end{center}
\caption{We plot for the case $U=0.69$ and $N=10000$ (using  M1
i.c.) the cumulative distribution of the clusters size, calculated
for three times $(t=200,350,500)$ in the QSS region. See text.}
\label{fig2}
\end{figure}

As required by the discovery of correlations and aging, which in
turn imply complex trajectories of the system in  phase space, it
is interesting to explore directly the microscopic evolution of
the QSS. This can be easily done plotting the time evolution of
the \textit{Boltzmann $\mu$-space}, where each particle of the
system is represented by a point in a plane characterized by the
conjugate variables $\theta_i and  p_i$, respectively the angular
position and the velocity of the $ith$ particle .
\\
It has been shown\cite{plu1} that, during the QSS regime,
correlations, structures and clusters formation in the $\mu$-space
appear for the $M1$ i.c., but not for initial conditions with zero
magnetization, the so called $M0$ i.c.: in the latter case both
the angles and velocities distributions remain homogeneous from the beginning
and a very slow mixing of the rotators has been observed.
For the $M1$ case, the dynamics in $\mu$-space can
be clarified through the concept of "dynamical frustration": the
clusters appearing and disappearing on the unit circle compete one with each other
in trapping more and more particles, thus generating a dynamically frustrated
situation that put the system in a glassy-like regime.
\\
In Fig.\ref{fig1} we show  a  molecular dynamics simulation
where the complete distribution function $f(\theta, p, t)$ is
considered for different values of time.
In fact, we plot - for M1 i.c., N=10000 and U=0.69 - a sequence of  
snapshots of $f(\theta, p, t)$ for six different times: at the
beginning of the simulation (t=0), in the QSS regime (t=50-500)
and towards canonical equilibrium (t=10000). In the QSS region one clearly
observes the presence of competing clusters, each cluster being composed by
particles with both angles and velocity included in the same
$\mu$-space cell (notice that, in our simulations, we considered a total of 100x100 cells for the $\mu$-space lattice). For t=10000, instead, any trace of macroscopic glassy behavior has disappeared.
\\
In Fig.\ref{fig2} we show the power law behavior of the cluster
size cumulative distributions calculated in the case $U=0.69$ for
several snapshots in the QSS regime at time t=200,350 and 500.
For each one of the 100x100 cells a sum over 20 different realizations (events)
has been performed. Then, for each cluster size (greater than
 5 particles) the sum of all the clusters with that size has been calculated and plotted.
As one can see from \ref{fig2}, the distribution does not change
significatively in the plateaux region as expected.
We report also a power law fit (drawn as a straight dashed
line above the data points) which 
indicates that the cluster distribution has an
approximately exponent decay $-1.6$.
The cluster size distribution reminds closely that of percolation
at the critical point, where a lenght scale, or time scale,
diverges leaving the system in a self-similar state
\cite{binney}.
More in general, it has been also suggested \cite{sotolongo} that,
optimizing Tsallis' entropy with natural constraints in a regime of long-range correlations, it is possible to derive a power-law hierarchical cluster size
distribution which can be considered as paradigmatic of physical
systems where multiscale interactions and
geometric (fractal) properties play a key role in the relaxation
behavior of the system.
Therefore, we can say that the power-law scaling resulting in the distributions of Fig.\ref{fig2} strongly suggests a non-ergodic topology of a region  of  phase
space in which the system remains trapped during the QSS
regime (for the M1 i.c.), thus supporting the weak-ergodicity
breaking scenario.
A weak breakdown of ergodicity, as originally proposed by Bouchaud
for glassy systems \cite{bouchaud}, occurs when the phase-space is
a-priori not broken into mutually inaccessible regions in which
local equilibrium may be achieved, as in the true
ergodicity-breaking case, but nevertheless the system can remain
trapped for very long times in some regions of the complex energy
landscape. In fact it is widely accepted that the energy landscape
of a finite disordered (or frustrated)
 system is extremely rough, with many local minima
 corresponding to metastable configurations.
Since the energy landscape
 is rough, these local minima are surrounded by rather high energy barriers and we
 thus expect that these states
 would act as "traps" which get hold of the system during
 a certain time $\tau$.
\\
In ref.\cite{bouchaud} such a mechanism has been proposed in order
to explain the aging phenomenon, i.e. the dependence of the
relaxation time on the history of the system, i.e. on its
\textit{age} $t_w$. Actually, it results that $\tau_{max} \simeq
t_w$, being $\tau_{max}$ the longest trapping time actually
encountered during a waiting time $t_w$. In other words, the
deepest state encountered traps the system during a time which is
 comparable to the overall waiting time, a result that - in turn - 
 allows to quantitatively  describe the relaxation laws
  observed in glassy systems \cite{spin-glass}.
\\
Aging phenomenon has been found also in the HMF model for $M1$
i.c., more precisely in the autocorrelation functions decay for
both the angles and velocities \cite{celia1} and for velocities
 only \cite{plu1}, thus reinforcing the hypothesis that a weak
 ergodicity-breaking could really
 occur in the metastable QSS regime and could be related to the
  complex dynamics generated
 by the vanishing of the largest Lyapunov exponent and by the
 dynamical frustration due to
the many different small clusters observed in this regime. Such a
scenario is in agreement also with the results about anomalous
diffusion shown in ref.\cite{hmf-dif}, where the probability
distribution of the trapping times, calculated for a test particle
in the transient QSS regime for $M1$ i.c., shows a clear power law
decay
\begin{equation}
    P_{trap}    \sim  t^{-\nu}~~,
\label{ptrap}
\end{equation}
characterized by an exponent $\nu$ related to the anomalous
diffusion coefficient.
In the next section we will show that the anomalous diffusion
coefficient can in turn be connected with the velocity
correlations decay by means of the nonextensive formalism, thus
suggesting a deeper link between the latter and the weak
ergodicity breaking.

\subsection{{Nonextensive thermodynamics and HMF model}}

In previous works it was  shown that the majority of the dynamical
anomalies of the QSS regime, among which $\mu$-space correlations,
clusters and dynamical frustration, are present not only for $M1$
initial conditions, but also when the initial magnetization
$M(t=0)$ is in the range $(0,1]$ \cite{plu3}.
\\
In order to prepare the initial magnetization in the range $0 < M \le 1$,
we distribute uniformly the particles into a variable portion of
the unitary circle. In this way we fix the initial potential
energy $V(\theta)$ and, in turn, the magnetization. Finally, we
assign the remaining part of the total energy as kinetic energy by
using a  water bag uniform distribution for the velocities.
The velocity correlations can be calculated by using the following
autocorrelation function\cite{plu3}
\begin{equation}
{C}(t)= \frac{1}{N} \sum_{j=1}^N {p_j(t) p_j(0)} ~~,
\label{nn_corr}
\end{equation}
where $p_j(t)$ is the velocity of the $j$-th particle at the time $t$.
\begin{figure} 
\label{fig3}
\begin{center}
\epsfig{figure=correlations.eps,width=6truecm,angle=0}
\epsfig{figure=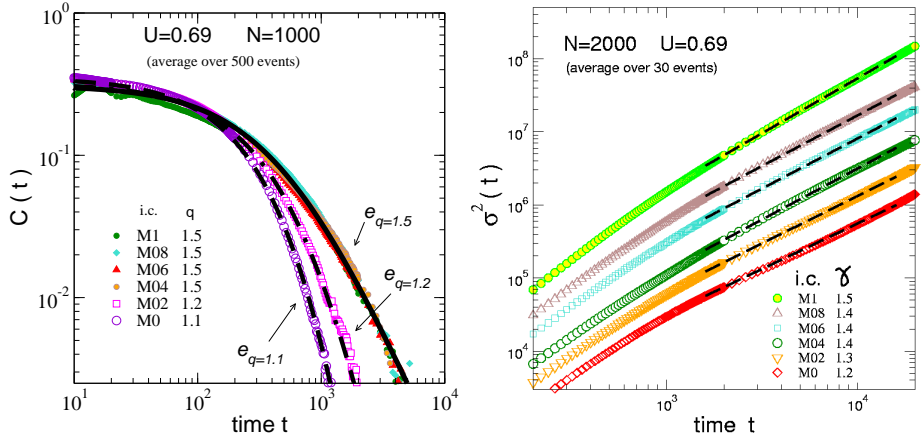,width=6truecm,angle=0}
\end{center}
\caption{(Left) Correlation functions vs time for different initial magnetization
(symbols). The solid lines are normalized q-exponentials.
(Right) We plot the mean square displacement of the angular motion
$\sigma^2 \propto t^\gamma$ vs time for  different initial
magnetizations. The exponent $\gamma$ which fit the data and charcterize the
behavior in the  QSS regime  is also
reported for the different initial conditions used. The dashed lines have  a slope corresponding   to these values.
}
\end{figure}
In Fig.\ref{fig3}-left, 
we plot the velocity autocorrelation function (\ref{nn_corr}) for
$N=1000$ and $M(0)={1, 0.8, 0.6, 0.4, 0.2, 0}$.
An ensemble average over 500 different realizations was  performed.
For $ M(0)\ge 0.4$ the correlation functions are very similar,
while the decay is faster for $M(0)=0.2$ and even more for $M(0)=0$.
If we fit these relaxation functions by means of the 
Tsallis' q-exponential function
\begin{equation}
e_q(z)= {\left[  1+(1-q) z \right]} ^\frac{1}{(1-q)}~~,
\end{equation}
with $z=-\frac{p}{\tau}$, and where $\tau$ is a characteristic
time, we can quantitatively discriminate between the  different
initial conditions. In fact we get a q-exponential with $q=1.5$
for $M(0)\ge0.4$, while we get $q=1.2$ and  $q=1.1$ for $M(0)=0.2$
and for $M(0)=0$ respectively. Notice that for $q=1$ one recovers
the usual exponential decay \cite{tsallis1,hmf2,hmf3,plu1}. Thus for
$M(0)>0$ correlations exhibit a long-range nature and a slow
power-law  decay. This decay is  very similar for $M(0)\ge 0.4$, but  
diminishes progressively below $M(0)=0.4$ to become almost exponential for $M(0)=0$.
\\
In order to study diffusion, one  can consider  the mean square displacement
of phases $\sigma^{2}(t)$ defined as
\begin{equation}
\sigma^{2}(t) = \frac{1}{N} \sum_{j=1}^{N}
  [ \theta_{j}(t) - \theta_{j}(0) ]^{2}
  = < [ \theta_{j}(t) - \theta_{j}(0) ]^{2} > .
\label{msd2}
\end{equation}
where the symbol $<...>$ represents the average over all the $N$
rotators. The quantity $\sigma^{2}(t)$ typically scales as
$\sigma^{2}(t)\sim t^{\gamma}$. The diffusion is normal when
$\gamma=1$ (corresponding to the Einstein's law for Brownian
motion) and ballistic for $\gamma=2$ (corresponding to free
particles). For different values of $\gamma $ the diffusion 
is anomalous, in particular for $1<\gamma<2$ one has superdiffusion. 
We  notice that the quantity $\sigma^{2}(t)$
can be rewritten by using the velocity correlation function $C(t)$
as
\begin{equation}
    \sigma^{2}(t)
     = \int_{0}^{t} dt_{1} \int_{0}^{t} dt_{2}~
    < p_{j}(t_2)~p_{j}(t_1) > \\
     = 2 \int_{0}^{t} dt_1 \int_{0}^{t_1} dt_{2}~
    C(t_2)~~~,
\label{sigma_corr}
\end{equation}
where $C(t)$ is defined as in Eq.\ref{nn_corr}.
\\
Superdiffusion has been already observed in the HMF
model  for  M1 initial conditions\cite{hmf-dif}. Rercently  we have also
checked that, even decreasing the initial magnetization, the
system continues to show superdiffusion \cite{plu3}. We illustrate
this behavior in Fig.\ref{fig3}-right,
where one sees taht, after an initial  ballistic
 regime  $(\gamma=2)$ proper of the initial  fast relaxation, the system shows 
 superdiffusion in correspondence of 
the QSS  plateau region and afterwards.
The exponent goes progressively from $\gamma=1.4-1.5$ for $0.4<M(0)<1$ to
 $\gamma= 1.2$
for M0. In the latter case, we have checked that, by  increasing the
size of the system,  diffusion tends to be normal ($\gamma\sim 1$ for
N=10000).
\begin{figure}[t]
\label{fig4}
\begin{center}
\epsfig{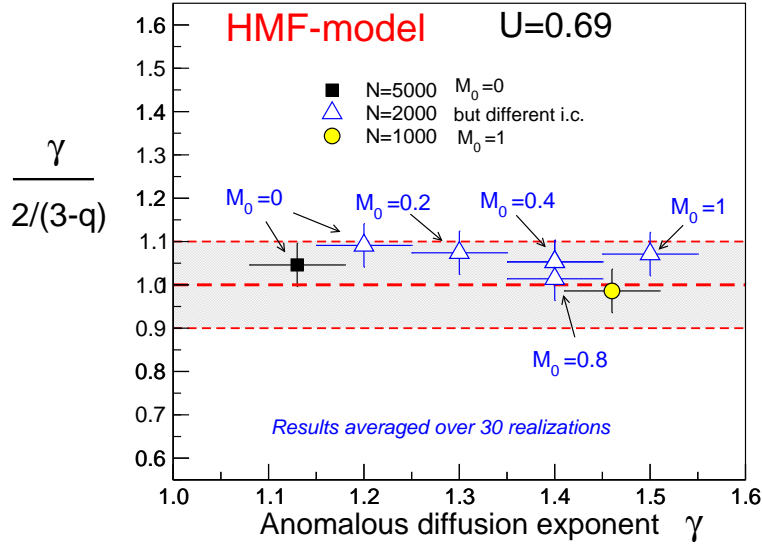}
\end{center}
\caption{In this figure we check the correctness of the $\gamma-q$
conjecture (see text) for
 the HMF model by plotting $\gamma$ as a function of the ratio $2/(3-q)$
 for different initial
  conditions with variable magnetization. See text.
}
\end{figure}
The slow decay and the superdiffusive behavior 
illustrated  in Fig.\ref{fig3} can be connected by
means of a conjecture based on a theoretical result found in 
ref.\cite{tsa-buk} by
Tsallis and Bukman. In fact in that paper the
authors show, on {\it general} grounds, that non-extensive
thermostatistics constitutes a theoretical framework within which
the {\it unification} of normal and {\it correlated} (driven)
anomalous diffusions can be achieved. They obtain, for a generic
linear force $F(x)$, the physically relevant {\it exact} (space,
time)-dependent solutions of a generalized Fokker-Planck (FP)
equation
\begin{equation}
\frac{\partial}{\partial t}[p(x,t)]^\mu=-\frac{\partial}{\partial x}
\left\{F(x)[p(x,t)]^\mu\right\}+D \frac{\partial^2}{\partial x^2}[p(x,t)]^\nu
\label{nlfpe}
\end{equation}
by means of an ansatz based on the Tsallis entropy.
\\
For our purpose, we remind here that such a FP equation, in the
nonlinear "norm conservation" case ($\nu \neq 1$ and $\mu=1$),
generates Tsallis space-time distributions with the entropic index
$q$ being related to the parameter $\nu$ by $q = 2 - \nu$. By
means of the latter, and following again ref.\cite{tsa-buk}, it is
possible to recover the following relation between the exponent
$\gamma$ of anomalous diffusion (being $\sigma^2 \propto
t^{\gamma}$) and the entropic index $q$
\begin{equation}
\gamma = \frac{2}{1+\nu} = \frac{2}{3-q}.
\label{gqrel}
\end{equation}
Hence, being in diffusive processes the space-time distributions
linked to the respective velocity correlations by the relation
(\ref{sigma_corr}), one could think to investigate if the relation
(\ref {gqrel}) would be satisfied choosing the entropic index $q$,
characterizing the correlation decay, and the corresponding  anomalous diffusion
exponent. 
This is done in Fig.4 where in order to check the latter hypothesis, that we call
 the {\it
$\gamma-q$ conjecture}, we report the ratio $\frac{\gamma}{2/(3-q)}$ vs the exponent 
$\gamma$ for various initial conditions ranging from
M(0)=1 to M(0)=0  and different sizes at $U=0.69$. Both $q$ and $\gamma$ have been taken from the results shown in Fig.\ref{fig3}.
Within an uncertainty of $\pm 0.1$, the data show that this  ratio is always one, thus providing a  strong 
indication in favor of this conjecture, which is satisfied for the  HMF model.
\\
Summarizing, we have shown  numerical simulations 
which connect the superdiffusion observed in the anomalous QSS regime of the
 HMF model  to  the $q$-exponential long-term decay of the
velocity correlations in the same regime.
 This new result is very interesting because opens
 a way to set a rigorous analytical link
 between the entropic
  index $q$ and the dynamical properties of nonextensive Hamiltonian
  many-body systems.

\section*{Conclusions}
We have briefly reviewed some of the anomalous features oberved in
the dynamics of the HMF model, a kind of minimal model
for the study of complex behavior in systems with long-range interactions.
We have also discussed how the anomalous behavior can be interpreted
within the nonextensive thermostatistics introduced by Tsallis,
and in the framework of the theory glassy systems.
The two pictures are not in contradiction and probably have more
strict links than previously thought, which
deserve to be further explored in the future.

\end{document}